\documentclass[a4paper,11pt]{article}
\usepackage{pos}
\usepackage{booktabs}

\title{Particle physics science communication with grass-roots Instagram: a success story}
\ShortTitle{Particle physics science communication with grass-roots Instagram}

\author*[a,b]{Freya Blekman}
\author[a]{Lucia Coll Saravia}

\affiliation[a]{Deutsches Elektronen-Synchrotron DESY, \\ Notkestr. 85, 22607 Hamburg, Germany}

\affiliation[b]{Institut f\"ur Experimentalphysik, Universit\"at Hamburg, \\
Luruper Chaussee 149, 22761 Hamburg, Germany}

\emailAdd{freya.blekman@desy.de}

\abstract{Communicating science to the public is an integral part of the work of a research collaboration, not an afterthought.
Institutional and collaboration-level social media accounts are often run by professional communicators rather than by practising scientists, and single-scientist ``influencer'' accounts are demanding to sustain.
We describe {\em CMSatDESY}, a crowd-sourced Instagram account on which the CMS group at DESY tells its own stories, each member in their own voice.
In the first half of 2026, the science communication activities through this reached 140k+ unique accounts and received 250k+ views, with an engagement rate of 7\%.
We discuss experiences and best practices that can be used by other groups, and describe how the account doubles as a science communication playground for outreach training for a large particle physics group.
}

\FullConference{43rd International Conference on High Energy Physics (ICHEP 2026)\\
30 July  to 5 August , 2026\\
Natal, Brazil\\}

\begin{document}
\maketitle

\section{Introduction}
In present society and funding conditions, communicating science is not an afterthought.
Public understanding of, and trust in, fundamental research underpins its funding and its place in society, and a well-communicated result is also more likely to be read, cited and built upon by colleagues.
Communication aimed at stakeholders outside the field, such as funding bodies, policy makers, students and the general public, relies on additional and a different balance of skills than communication within the field, and these are skills that scientists are usually not trained for.

In that context, we present {\em CMSatDESY} \cite{Blekman:2024qig}, an Instagram account run by the CMS~\cite{CMS:2008xjf} group at DESY, and active since 2024. 
DESY is a German research centre with about 2700 staff and a further 500 PhD students and postdoctoral researchers.
CMS is one of the largest experiments at CERN, with about 4000 scientists from 240 institutes.
The CMS DESY group consist of 80+ scientists varying from bacheler studunts to emeriti, and is active in detector hardware, physics analysis and data taking, among other activities. While the entire group contributes content, the daily operation of the instagram account is run by a small group of scientists that are also active in research and contribute a small fraction of their time to the management/organisation. 

\section{Why not use institute or collaboration social media?}
Social media is crucial for bridging the gap between scientists and the public, and also between different physics disciplines.
Contributing to it teaches scientists to convey complex ideas in a relatable manner, which in turn fosters trust and engagement from funders and from society, and enhances public understanding of science.

The official accounts of the laboratory, institute or collaboration are often run by communication professionals rather than by practising scientists.
This has clear advantages: the staff are trained in communication and their output is often excellent.
It also carries a risk, particularly on social media.
If it is not explicit that the person speaking is a practising scientist, the account can be perceived as the voice of a large institution rather than that of a researcher.
Audiences, particularly science skeptics and deniers, notice this and push back, particularly against communication that appears insincere or corporate, and against ``fake experts''.
In this context ``fake'' can simply mean someone who represents anonymous institutional monolithic science and appears to claim scientist status without actually being a scientist. The problem is aggravated by the growing volume of science misinformation on social media. The perceived trustworthiness and credibility of science on social media when scientists are presented as disingenuous is well studied~\cite{Gierth2020} and by now science communication by working scientists is strongly encouraged by various sources including the funding agencies in Germany~\cite{naturescicomm,bmftr,Chiarelli:2022tpd}.

\section{Why not a single person?}
The obvious alternative to institutional communication is a single scientist who builds a personal presence on a platform; for scientists communicating in new media environments in general and in particle physics see Ref.~\cite{Liang2014,cerncouriertiktok}.
We are generally enthusiastic about this model.
A parasocial bond with the audience makes science human, and it puts a practising scientist at the centre of the communication.

However, scientists working as high profile influencers create several drawbacks:
\begin{itemize}
\item It requires a great deal of hard, intensive work that is not easily combined with an already very time-consuming science career.
\item Followers are hard to build up, so the first years reach very few people.
\item Many scientists may not be interested in investing the time in these efforts.
\item It denies the broader community the opportunity to build these essential skills that are also used for other aspects of science careers.
\end{itemize}

\section{A crowd-sourced approach: {\em CMSatDESY}}
Our solution is an Instagram account that carries all of our stories, told many different members of the group.
The workload is shared, the voice is that of practising scientists, and the account always shows a person rather than an institution. Because the work is shared and individual scientists only contribute occasionally, the work load stays low for individuals while still reaching substantial impacts. Contributors do not have to handle the platform side.
Editing, captions, hashtags and scheduling are done by a small team that runs the account.
This team helps and guides their colleagues from raw material: an idea (a recent result, a project, or a slice of a day that an outsider might find interesting), a small amount of raw material (a few short video clips, or a handful of photos, or a paragraph of text; phone quality is sufficient), and a conversation with a team member to shape the post. Most post are made by contributors after that, and contributors spend one to two hours in total to prepare a post.

To motivate scientists, contributors are provided four reasons to post: outreach, science communication, training and recruitment.
The content policy is deliberately light: \emph{everything is allowed as long as you do not embarrass us or yourself}. The efforts purposefully highlight all career levels and do not follow a fixed format, and creativity is encouraged, providing a clear ``made by a scientist not a PR agency'' visual style. 
This includes directly following and contributing to memes and trending content, which may be too frivolous or not newsworthy enough for central lab communications.
An important recurring theme is focusing on the personal experience of what it is like to be a scientist, ``Our journey to particle physics'': contributors talk about the path that brought them to the field, reflect on unique experiences, or simply talk about their PhD.

\section{Performance}
Table~\ref{tab:numbers} summarises the account statistics for 1 January to 28 July 2026.
Reach is the number of unique accounts, views are video plays, and the engagement rate is defined as interactions divided by views.
The per-post values are consistent with the totals for 58 posts.
Typical follower growth is 5\% per month, which is a competitive performance that is high in the guidelines provided by CERN for social media success.

\begin{table}[ht]
\centering
\begin{tabular}{lr}
\toprule
Quantity (1 Jan -- 28 Jul 2026) & Value\\
\midrule
Posts & 58\\
Reach (unique accounts) & 142\,210\\
Views (video plays) & 253\,948\\
Engagement rate (interactions $\div$ views) & 6.99\%\\
Reach per post & 2452\\
Views per post & 4378\\
Likes per post & 137\\
Typical follower growth per month & 5\%\\
\bottomrule
\end{tabular}
\caption{Key statistics of the {\em CMSatDESY}\ account in the first half of 2026.}
\label{tab:numbers}
\end{table}

\subsection{Audiences reached}
The age distribution of followers for the {\em CMSatDESY} account is typical for Instagram, with 25--34 year olds the largest group (41\%) followed by 18--24 year olds (27\%); the 13--17 group accounts for 1\% so this account reaches different audiences as outreach activities at schools and the International Particle Physics Masterclasses~\cite{IMCS}
The audience is highly international as this account is operated in English. Germany (17\%), India (13\%), the United States (11\%) and Italy (5\%) are the largest individual countries, while 54\% of followers are elsewhere.
Women make up 32\% of followers and men 68\%, which is a reasonable gender diversity for a physics account.

\subsection{Good practices}

As this instagram account is by now reaching substantial audiences, we have identified practices that can be useful for others who are interested in a similar effort. It is useful to state that many of these depend on the propriatory instagram algorithmic settings that may change over time, but in general these will likely remain valid for a while.

\paragraph{Technical topics can do well, if explained at the right level.}
Technical topics perform very well, as the technology and link to impressive scientific machines can be very inspiring to the audience. It is important to tailor the explainations to the right level, but if that consideration is respected, graphical explanations of detector construction reached tens of thousands of people in a single post.

\paragraph{Personal stories work.}
The personal ``journey to particle physics'' videos and slide shows (known as carrousels for instagram) perform well. The personal stories reach large audiences, reach people outside the existing follower base, and prompt likes, comments and saves. 

\paragraph{Mixed language works.}
We automatically translate reels into the languages of the largest follower groups, for example German, major Indian languages and Spanish, and find that mixed-language content successful.

\section{A second objective: training}
Beyond reaching the public, a second objective of {\em CMSatDESY}\ is to provide the entire group of more than 80 people with basic training on how to communicate science effectively.
The training rests on four principles that are also the foundations of stakeholder communications training, but presented at the level that scientists can engage with it without prior media training:
\begin{description}
\item[Audience] put yourself in the perspective of your audience;
\item[Lead] start with a clear purpose and message;
\item[No jargon] simplify technical terms for better comprehension;
\item[Storytelling] engage through personal stories and analogies rather than just facts.
\end{description}
Overall, the instructions to contributors are founded on modern science-communication planning principles, which we try to instil in everyone involved: identify the stakeholders, define the communication objectives, select the channels, develop a schedule, execute the plan, and monitor and always evaluate its effectiveness.
The Instagram insights provide this monitoring directly, but importantly we also provide this information back to the scientist-contributors and their line managers, in such a way that this can potentially be used as an outreach contribution item on academic CVs.

\section{What contributors gain}
Contributors benefit in several ways:
\begin{itemize}
\item They explain their own work from the outside in, which provides training in science communication at a level comfortable for each person.
\item They receive immediate feedback from a non-expert audience through the Instagram statistics, which are more informative than one might expect.
\item They obtain a public, timestamped record of the physics they worked on, and posts that perform well can be added to an academic CV (``I made an Instagram video about my thesis that was viewed by 50\,000 people'').
\item They have a low-stakes space in which to experiment with communicating their work before it matters.
\item The account is also a recruitment tool: a large fraction of prospective MSc and BSc students, and summer students, already know our Instagram account before they apply for a research project in our group.
\end{itemize}

\section{Conclusions}
A crowd-sourced, practitioner-led Instagram account is a complementary communication efforts to institutional accounts and to single-scientist accounts, with their heavy workload.
In the first seven months of 2026, {\em CMSatDESY}\ reached more than 140\,000 unique accounts, while giving an 80+ person group direct practical experience of communicating their work.

\section*{Acknowledgements}
The authors acknowledge support from DESY (Hamburg, Germany), a member of the Helmholtz Association HGF, and support by the Deutsche Forschungsgemeinschaft (DFG, German Research Foundation) under Germany’s Excellence Strategy -- EXC 2121 ``Quantum Universe'' -- 390833306.

\end{document}